\documentclass[aps,nofootinbib,11pt,preprintnumbers]{revtex4}
\usepackage{amssymb,amsmath}
\usepackage{hyperref}
\usepackage[dvips]{graphicx}
\newcommand{\bea}{\begin{eqnarray}}
\newcommand{\eea}{\end{eqnarray}}

\def\x{{\bf x}}

\begin{document}
\preprint{MIT-CTP 4014}

\title{
How many phases meet at the chiral critical point?
}
\author{Dominik Nickel}
\affiliation{Center for Theoretical Physics,
  Massachusetts Institute of Technology,
  Cambridge, MA 02139, USA}
\date{February 2009}

\begin{abstract}
\noindent
We explore the phase diagram of NJL-type models near the chiral critical point allowing for phases with spatially inhomogeneous chiral condensates.
In the chiral limit it turns out that the region in the mean-field phase diagram where those phases are energetically preferred very generically reaches out to the chiral critical point.
The preferred inhomogeneous ground state in this vicinity possibly resembles a lattice of domain wall solitons.
This raises the question of their relevance for the phase diagram of QCD.
\end{abstract}

\maketitle

\section{Introduction}
\noindent
The phase diagram of quantum chromodynamics (QCD) as a function of temperature $T$ and quark chemical potential $\mu$ is expected to exhibit a rich phase structure~\cite{reviews}.
In particular the nature of the chiral phase transition, the location of the chiral critical point and the properties in its vicinity have attracted a lot of interest.
Experimentally, this exploration is one of the main research goals at the future FAIR facility in Darmstadt~\cite{Senger:2006zz}.
Theoretically, it is heavily investigated in phenomenological models whereas ab initio calculations are still limited to small values in $\mu/T$.
Depending on the approach and/or details of the model a great range for the possible location of the critical point has been found~\cite{Stephanov:2007fk}.

Generic, model-independent scenarios near the critical point can however in principle be discussed via a Ginzburg-Landau (GL) expansion. The idea is to expand the effective action of the order parameter and to obtain an expression for the thermodynamic potential as a functional in the order parameter.
For inhomogeneous phases, characterized by a spatially varying order parameter, this can be generalized by a combined gradient expansion.
Treating gradients and the magnitude of the order parameter to be of the same order, a generic expression for the GL functional in the vicinity of a second order phase transition for a theory with real order parameter $\phi(\x)$ invariant under $\phi(\x)\rightarrow -\phi(\x)$ takes the form
\begin{widetext}
\bea
\label{eq:GLintro}
\Omega_{GL}(T,\mu;\phi(\x))
&=&
c_2(T,\mu)\phi(\x)^2
+
c_{4,a}(T,\mu)\phi(\x)^4
+
c_{4,b}(T,\mu)(\nabla\phi(\x))^2
\nonumber\\&&
+
c_{6,a}(T,\mu)\phi(\x)^6
+
c_{6,b}(T,\mu)(\nabla\phi(\x))^2\phi(\x)^2
+
c_{6,c}(T,\mu)(\Delta\phi(\x))^2
\,.
\eea
\end{widetext}
Considering a periodic ground state with a Wigner-Seitz cell $V$, the thermodynamic potential difference between symmetric and spontaneously broken phase is then given by $\Delta\Omega=\frac{1}{V}\int_V \Omega_{GL}$. Furthermore we have to require $c_{6,i}>0$ in order to have a bounded potential.
For homogeneous phases with $\phi(\x)=\phi_0$ a second order phase transition is then occuring at $c_{2}=0$, $c_{4,a}>0$, which turns into a weak first order phase transition at the critical point where $c_{4,a}$ changes sign. 
In addition inhomogeneous phases can become energetically favored when $c_{4,b}<0$, since small curvatures then lead to a gain in free energy.

The main idea of this work is to perform such an expansion within an NJL-type model on mean-field.
For simplicity we limit ourself to the chiral limit and do not include $U_A(1)$-breaking terms.
Furthermore constraining ourself to condensates of the form $\langle\bar{\psi}_i(x)\psi_j(x)\rangle\propto\phi(\x)\delta_{ij}$, where $i,j$ label flavors and chiralities, the residual global symmetry is $\mathbb{Z}_2$ and the GL functional should take the form as stated in Eq.(\ref{eq:GLintro}).
Assuming a sensible regularization procedure we find $c_{4,a}=c_{4,b}$ and related to that an inhomogeneous phase starting at the critical point.
 
This finding is actually very similar as in the one-dimensional\footnote{In this work we only refer to spatial dimensions.} Gross-Neveu (GN) model, which is renormalizable and where the mean-field problem can be solved analytically~\cite{Schnetz:2004vr,Thies:2006ti}.
In the mean-field phase diagram of the GN model a symmetric, a homogeneous dynamically broken and a inhomogeneous dynamically broken phase exist and meet at the chiral critical point.
A similar behavior persists also for finite quark masses ~\cite{Boehmer:2007ea}.

To conclude the introduction we want to clarify, that we do not necessarily want to conjecture precisely this scenario for the phase diagram of QCD.
Rather we would like to suggest that inhomogeneous phases may play a more important role than anticipated.
It should be noted that the NJL model shares the global symmetries with QCD and is therefore a reasonable candidate to study the phase properties near the critical point even if the location of the critical point may differ.
So far these chiral crystalline phases have mainly been discussed at vanishing temperatures~\cite{Deryagin:1992rw,Shuster:1999tn,Park:1999bz,Rapp:2000zd}.
Here the discussed scenarios are qualitatively different in that the modulations, i.e. the magnitude of the gradients, are not small but typically of the order of the Fermi momentum.
The relevant dynamics of particle-hole scattering for one plane wave is then concentrated on a resticted vicinity of two antipodal points on the Fermi surface.
An exception to this is the dual chiral-density wave which is sufficiently simple to solve the mean-field problem~\cite{Nakano:2004cd}. On a technical level it has strong similarities with the Fulde-Ferrell phase~\cite{Fulde:1964zz,Alford:2000ze} discussed in the context of (color-)superconductors.
The latter is known to be technically simple but disfavored compared to other inhomogeneous phases~\cite{LO64,Bowers:2002xr}.
In particular the obtained order of phase transitions when going from homogeneous dynamically broken to inhomogeneous to unbroken phase might be misleading~\cite{Nickel:2008ng}.

\section{The generalized Ginzburg-Landau expansion}
\noindent
We consider NJL-type models of the form
\bea
\mathcal{L}
&=&
\bar{\psi}
i\gamma^\mu \partial_\mu
\psi
+
G(\bar{\psi}\psi)^2
+
\dots
\,,
\eea
where $\psi$ is the $4N_f N_c$-dimensional quark spinor for $N_f$ flavors and $N_c$ colors, $\gamma^\mu$ are Dirac matrices and $G$ is called the scalar coupling.
The dots indicate terms that vanish in our mean-field approximation.
Allowing a mean-field value $\langle\bar{\psi}(\x)\psi(\x)\rangle = -\frac{1}{2G}M(\x)$ for the chiral condensate, the mean-field Lagrangian density takes the form
\bea
\mathcal{L}_{MF}
&=&
\bar{\psi}(i\gamma^\mu \partial_\mu - M(\x))\psi -\frac{M(\x)^2}{4G}
\,.
\eea
In the case of a periodic condensate with Wigner-Seitz cell $V$ and using the imaginary-time formalism (see e.g. Refs.~\cite{reviews}), we therefore obtain for the mean-field thermodynamic potential
\begin{widetext}
\bea
\label{eq:Omega1}
\Omega(T,\mu)
&=&
-\frac{T}{V}\ln 
\int \mathcal{D}\bar{\psi}\mathcal{D}\psi \exp\left(\int_{x\in [0,\frac{1}{T}]\times V} (\mathcal{L}_{MF}+\mu \bar{\psi}\gamma^0 \psi)\right)
\nonumber\\
&=&
-\frac{T}{V}
\mathrm{Tr}_{D,c,f,V} \, \mathrm{Log}\left(S^{-1}\right)
+
\frac{1}{V}\int_V
\frac{M(\x)^2}{4G}
+\text{const.}
\,,
\eea
\end{widetext}
with inverse propagator
\bea
S^{-1}(x,y)
&=&
(i\gamma^\mu \partial_\mu - M(\x))\delta^{(4)}(x-y)
\eea
and the functional trace acting on the direct product of Dirac, color, flavor and coordinate space.
Since the functional logarithm can only be evaluated in special cases such as the
dual chiral-density wave~\cite{Nakano:2004cd}, we now expand in $M(\x)$. Substracting the leading order corresponding to the thermodynamic potential of the unbroken phase, we formally arrive at
\bea
\Delta\Omega(T,\mu)
&=&
-\frac{T}{V}
\sum_{n>0}\frac{1}{n}\mathrm{Tr}_{D,c,f,V}\left(S_0 M\right)^n
+
\frac{1}{V}\int_V
\frac{M(\x)^2}{4G}
\,.
\eea
Here we have introduced the bare propagator $S_0=S\vert_{M(\x)=0}$ and  a short hand notation for
\begin{widetext}
\bea
\mathrm{Tr}_{D,c,f,V}\left(S_0 M\right)^n
&=&
\int_{x}
\int_{x_2}
\dots
\int_{x_n}
\mathrm{Tr}_{D,c,f}
\left(
M(\x)
S_0(x,x_2)
M(\x_2)
\dots
M(\x_n)
S_0(x_n,x)
\right)
\,.
\nonumber\\
\eea
\end{widetext}
The domain of integration for $x$ is $[0,\frac{1}{T}]\times V$ and $[0,\frac{1}{T}]\times\mathbb{R}^3$ for $x_2,\dots,x_n$.
In the chiral limit the expressions for odd values of $n$ vanish. Furthermore we can expand the condensate around $\x$ as\footnote{We are using multi-index notation.}
\bea
M(\x_n)
&=&
\sum_{\vert\alpha\vert>0}\frac{1}{\alpha!}D^{\alpha}M(\x)(\x_n-\x)^{\alpha}
\eea
and can extract the GL functional to any desired order in gradients and order parameter.
Neglecting possible issues with the regularization for the moment, we can go to momentum space using
$
S_0(x,y)
=
T\sum_n \frac{d^3 p}{(2\pi)^{3}} ({p_\mu \gamma^\mu})^{-1}\exp(ip(x-y))
\,,
$
where $p_0 = \omega_n = (2n+1)\pi T$.
It is then a tedious but straightforward exercise to work out the explicit expression stated in Eq.(\ref{eq:GLintro}).
All coefficients result in the evaluation of similar integrals, only
\bea
c_{6,b}
&=&
- 32 N_f N_c G^4 
T\sum_n
\int \!\! \frac{d^3 p}{(2\pi)^{3}}\,\,
\left(
\frac{5}{3}\frac{1}{((\omega_n+ i\mu)^2+p^2)^{3}}
+
\frac{11}{18}\nabla_{\bf{p}}\cdot\frac{\bf{p}}{((\omega_n+ i\mu)^2+p^2)^{3}}
\right)
\eea
takes a slightly more complicated form, in that the integrand involves total derivatives.
In addition part of the expressions are of course formally divergent.

Since our model is non-renormalizable the divergences cannot be absorbed into a redefinition of the coupling. Instead usually a regularization, as part of the phenomenological model, is introduced.
Due to this ad hoc procedure a generalization of the regularization to inhomogeneous phases is often not unique\footnote{This applies in particular for the widely used three-momentum cutoffs/formfactors.} as already discussed in the context of inhomogeneous color-superconductors~\cite{Nickel:2008ng}.
We could therefore take the pragmatic viewpoint that a generalization of any such ad hoc regularization procedure to inhomogeneous phases is assumed to be such that the total derivative terms vanish.
An alternative approach is a regularization scheme that does not rely on an homogeneous ground state, e.g. a propertime regularization for the functional logarithm in Eq.(\ref{eq:Omega1}).
In this case it is possible to show that the total derivative terms strictly vanish.
For both of these viewpoints we can define the regularized expressions
\bea
\alpha_n
&=&
(-1)^{\frac{n}{2}} 4 N_f N_c 
T\sum_m
\int_{reg.} \!\! \frac{d^3 p}{(2\pi)^{3}}\,\,
\frac{1}{((\omega_m+ i\mu)^2+p^2)^{\frac{n}{2}}}
+
\frac{\delta_{2n}}{2G}
\eea
and (not replacing $M(\x)$ by $\phi(\x)$ for simplicity) finally obtain
\bea
\label{eq:OmegaGL}
\Omega_{GL}(T,\mu;M(\x))
&=&
\frac{\alpha_2}{2}
M(\x)^2
+
\frac{\alpha_4}{4}
\left(M(\x)^4 + (\nabla M(\x))^2\right)
\nonumber\\&&
+
\frac{\alpha_6}{6}
\left(
M(\x)^6
+
5(\nabla M(\x))^2 M(\x)^2
+
\frac{1}{2}(\Delta M(\x))^2
\right)
\,.
\nonumber\\
\eea
Interestingly the ratios of the prefactors within each order of the GL expansion turn out as those in the one-dimensional analogue~\cite{Schnetz:2004vr,Thies:2006ti}. This is not the case in superconductors~\cite{Buzdin:1997a} where the underlying dynamics is different, namely coming from particle-particle and hole-hole scattering near the Fermi surface instead of particle-hole scattering in presented case.

It is now in principle straightforward to choose a regularization scheme and adjust the model parameters to QCD phenomenology. Since it is however known that differing choices of regularizations and model parameters lead to a spread in the location of the critical point (see e.g. the collection in Ref.~\cite{Stephanov:2007fk}), we prefer to make a more qualitative statement and keep the discussion on the level of the GL coefficients $\alpha_n$\footnote{For all conventional choices the presented model has a critical point in the phase diagram where $\alpha_2=\alpha_4=0$ and $\alpha_6>0$.}.
Although the model may not have the same  critical point as QCD, it should be in the same universality class and therefore show a similar behavior in the vicinity of the critical point.

\section{Inhomogeneous ground states}
\begin{figure}[h]
\begin{center}
\includegraphics[width=\linewidth]{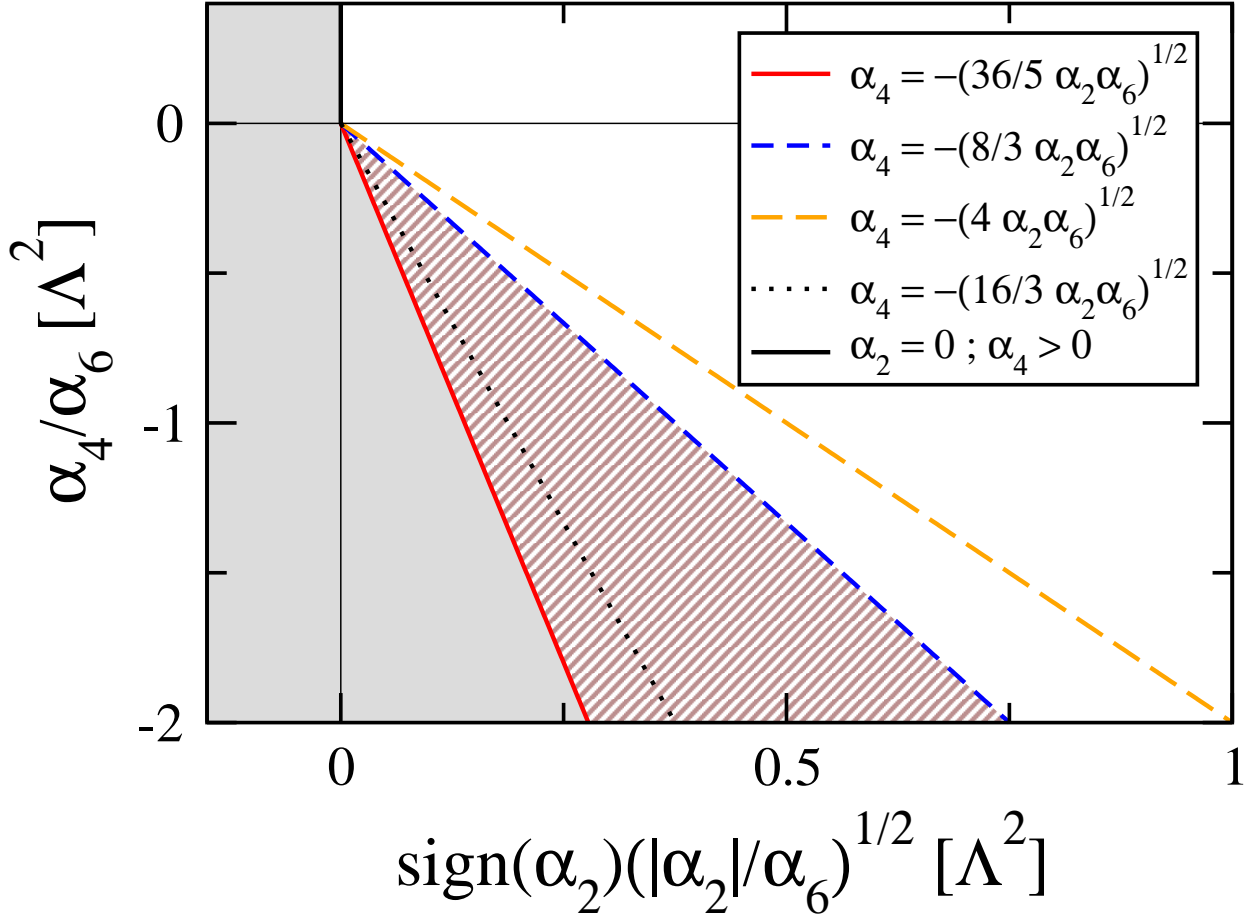}
\caption{
Pictorial presentation of the phase diagram in terms of the GL coefficients: The gray domain corresponds to the homogeneous dynamically broken ground state, the shaded gray to the solitonic ground state (at least when restricting to one-dimensional modulations in the order parameter), whereas in the transparent domain the unbroken phase is preferred. $\Lambda$ is an arbitrary scale. Also stated are various lines discussed in the text.
}
\label{fig:fig1}
\end{center}
\end{figure}

\noindent
Given the GL functional in Eq.(\ref{eq:OmegaGL}) and some critical point  where $\alpha_2 =\alpha_4=0$ and $\alpha_6>0$, we can explore the phase structure in its vicinity.
Limiting ourself first to homogeneous phases with $M(\x)=M_0$, we have the following known behavior:
\begin{description}
\item[$\alpha_4 > 0$:]
Second order phase transition at $\alpha_2 = 0$, where $M_0^2\simeq -\frac{\alpha_2}{\alpha_4}$ and $\Delta\Omega = -\frac{\alpha_2^2}{4\alpha_4}$ for $\alpha_ 2 < 0$.
\item[$\alpha_4 < 0$:]
First order phase transition at $\alpha_4 = -\sqrt{\frac{16}{3}\alpha_2 \alpha_6}$ where the broken solution has $M_0^2 = -\frac{3}{4}\frac{\alpha_4}{\alpha_6}$.
The dynamically broken solution continues to exist as a local minimum up to $\alpha_4 =-\sqrt{ 4\alpha_2 \alpha_6}$.
\end{description}
Allowing for inhomogeneous phases we may expect a spatially varying order parameter for $\alpha_4 <0$ since here small curvatures can lead to a gain in free-energy. Even within the GL approximation the determination of the ground state is not straightforward as we need to minimize a non-quadratic functional.
Focusing on one-dimensional inhomogeneities, i.e. $M(\x)=M(z)$, the solutions to $\frac{\delta}{\delta M}\Delta\Omega = 0$ are actually known from the investigation of one-dimensional models~\cite{Buzdin:1997a,Basar:2008im}. They are expressed (up to an arbitrary shift) in terms of the elliptic Jacobi $\mathrm{sn}$-function as
\bea
M_{1D}(z)
&=&
\sqrt{\nu} q \, \mathrm{sn}(q z,\nu)
\,,
\eea
where $\nu\in[0,1]$ and $q$ being a scale related to the maximum of $M_{1D}(z)$ and the extension of a soliton in the chosen $z$ direction (both scales are related in our case).
For $\nu=1$ we have $M_{1D}(z) = q \tanh(q x)$, i.e. a single soliton and for $\nu\rightarrow 0$ the shape becomes more and more sinusoidal albeit the amplitude also goes to zero.
From previous investigations~\cite{Schnetz:2004vr} it is known that when increasing $\alpha_2$ from zero we reach a second order phase transition into an inhomogeneous phase with $q=M_0$ and $\nu=1$. At this point the free-energy of a single soliton becomes negative leading to its formation.
By using $M_0$ known from above and checking where $\frac{d}{d\nu}\Delta\Omega\vert_{M(\x)=M_{1D}(z)}$ changes sign at $\nu = 0$, we obtain $\alpha_4 = -\sqrt{\frac{36}{5}\alpha_2 \alpha_6}$ for this point.
We arrive at the onset of infinitely far separated solitons. Further increasing $\alpha_2$ decreases $\nu$ until it reaches zero. Since $q$ stays finite the overall magnitude of $M_{1D}(z)$ given by $\sqrt{\nu}q$ then vanishes and we find a second order phase transition to the unbroken phase.

In case of a second order phase transition from the inhomogeneous phase to the unbroken phase, the value of $\alpha_4$ in terms of $\alpha_2 \alpha_6$ is actually universal also for higher dimensional modulations of the order parameter.
Since in this case $M(\x)$ is parametrically small, we can neglect non-quadratic terms in the GL functional.
Consequently the variation $\frac{d}{d M}\Delta\Omega$  leads to a linear partial differential.
We can then optimize the value of $\alpha_4$ by varying the momentum ${\bf{q}}$ of the Fourier components of $M(\x)$ and find $\alpha_4 =-\sqrt{ \frac{8}{3}\alpha_2 \alpha_6}$ for the transition line where $\vert{\bf{q}}\vert = \sqrt{-\frac{3\alpha_4}{2\alpha_6}}$.

We do not want to address the general question whether an inhomogeneous phase with a higher dimensional modulation could become favored in the vicinity of the critical points, but it may very well be that the one-dimensional modulations are generally preferred there as numerically confirmed in Ref.~\cite{Houzet:1999a} for the analogous case of inhomogeneous phases in paramagnetic superconductors.

We summarize our findings in Fig.~\ref{fig:fig1} by illustrating the ground states in the GL coefficients' phase diagram.
Obviously the inhomogeneous phases modify the region where a first oder transition is expected when restricting to homogeneous ground states only and replaces this transition by two second order phase transitions.

\section{Discussion}
\noindent
Starting with generic NJL-type models in mean-field approximation, we have preformed a generalized GL analysis in the vicinity of the chiral critical point. Since the order parameter and the gradients of the modulation are parametrically small, the expansion is arbitrary close to the mean-field description.
Requiring a sensible regularization scheme the results are astonishingly similar to results in the one-dimensional GN model.
Consequently, we find two second order phase transitions from homogeneous to inhomogeneous dynamically broken to unbroken phase, instead of a first order transition from broken to unbroken phase.

Various questions can still be addressed: It would be interesting to actually compute the GL coefficients in a regularization scheme with parameters adapted to QCD phenomenology and to estimate the extension of inhomogeneous phases in the phase diagram of NJL-type models. This should be combined by a complete mean-field study when at least allowing for a one-dimensional inhomogeneity in order to check the range for the applicability of the GL expansion. It should also give a different phase diagram than obtained for the analogous one-dimensional models, where the inhomogeneous phases reach out to infinitely high chemical potentials.
Another interesting question here is whether the order of the phase transitions change when going away from the critical point as recently found at vanishing temperatures for (color-) superconductors~\cite{Nickel:2008ng}.
Also the competition with color-superconducting phases would be interesting as the latter, at least in such model studies, should not reach out to the critical point.

There are also various ways the GL analysis can be improved: It would be interesting to include finite and mutually differing current quark masses and to check how the presented scenario is affected.
This would in addition require the use more complex order parameters $\langle\bar{\psi}_{i}\psi_{j}\rangle$, where the indices $i$, $j$ should include flavor and chirality.
Also more complicated interactions or models could be considered, such as vector interactions, the t'Hooft interaction when in particular including the strange quark or the Polyakov-loop NJL model.
Incorporating time derivatives one may try to explore dynamical properties of fluctuations and discuss possible consequences for heavy ion collisions.

More complicated but very interesting problems would be whether those phases could be explored beyond mean-field approximation, how the relation between the GL coefficients change when including higher fluctuations and whether the phase diagram of QCD may include those or similar phases, in particular in the region close to the chiral critical point. 

\acknowledgments
\noindent
We thank Michael Buballa, Gerald Dunne and Krishna Rajagopal for helpful comments.
This work was supported in part
by funds provided by the U.S. Department of Energy
(D.O.E.) under cooperative research agreement DE-FG0205ER41360 and
by the German Research Foundation (DFG) under grant number Ni 1191/1-1.


\end{document}